\documentclass[twocolumn,prl,aps,flushbottom,showpacs,10pt]{revtex4-1}
\usepackage{xspace,amsmath,amsfonts,amsthm,amssymb,amsbsy,graphicx,color}

\begin{document}

\newcommand{\nn}{\nonumber}
\newcommand{\ms}[1]{\mbox{\scriptsize #1}}
\newcommand{\dg}{^\dagger}
\newcommand{\smallfrac}[2]{\mbox{$\frac{#1}{#2}$}}
\newcommand{\la}{\langle}
\newcommand{\ra}{\rangle}
\newcommand{\ket}[1]{| {#1} \ra}
\newcommand{\bra}[1]{\la {#1} |}
\newcommand{\pfpx}[2]{\frac{\partial #1}{\partial #2}}
\newcommand{\dfdx}[2]{\frac{d #1}{d #2}}
\newcommand{\ioh}{-\frac{i}{\hbar}}
\newcommand{\ohh}{-\frac{1}{\hbar^2}}
\newcommand{\half}{\smallfrac{1}{2}}

\newtheorem{theo}{Theorem} 
\newtheorem{lemma}{Lemma}

\title{Twenty open problems in quantum control} 

\author{Kurt Jacobs}

\affiliation{ Department of Physics, University of Massachusetts at Boston, Boston, MA 02125, USA} 

\begin{abstract} 
The subject of controlling quantum systems is not new, but concepts that have been introduced in the last decade and a half, especially that of coherent feedback, suggest new questions that broaden and deepen the field. Here we provide a concise overview and definition of quantum feedback control, both coherent and measurement-based, and discuss its relationship to ``standard'' time-dependent control; there is a sense in which the latter subsumes the rest. There are many open questions within quantum control and its subfields, and we highlight and discuss some of them here. These questions are of theoretical as well as practical interest: the answers will help to determine the relative power of the different methods of control, and the limits to our ability to control quantum systems imposed by available resources.  
\end{abstract} 


\maketitle 

\section{Introduction: \\ feedback and open-loop control} 

When one thinks of feedback control, it is likely that the use of measurements will spring to mind. Measurements do not sit easily with quantum theory, and in fact they are somewhat superfluous. Measurements can always be replaced by considering instead the joint evolution of two systems, a primary system and an auxiliary. A measurement process on the primary constitutes the formation of correlations between the two, and feedback is simply the exploitation of these correlations by the joint evolution. In fact it is becoming increasingly clear that the ability to swap between a picture involving explicit measurements and one without measurements is useful. It suggests two alternative ways to physically implement the same control process, one in which measurement results are processed by classical computers, and one that merely involves the application of time-dependent control to a collection of quantum systems. While these two implementations are in a sense equivalent, they are very different from a practical point of view, and both have relative advantages. Further, while coherent feedback subsumes measurement-based feedback, the reverse is not true~\cite{Jacobs13}. 

\begin{figure}[t]
\leavevmode\includegraphics[width=0.85\hsize]{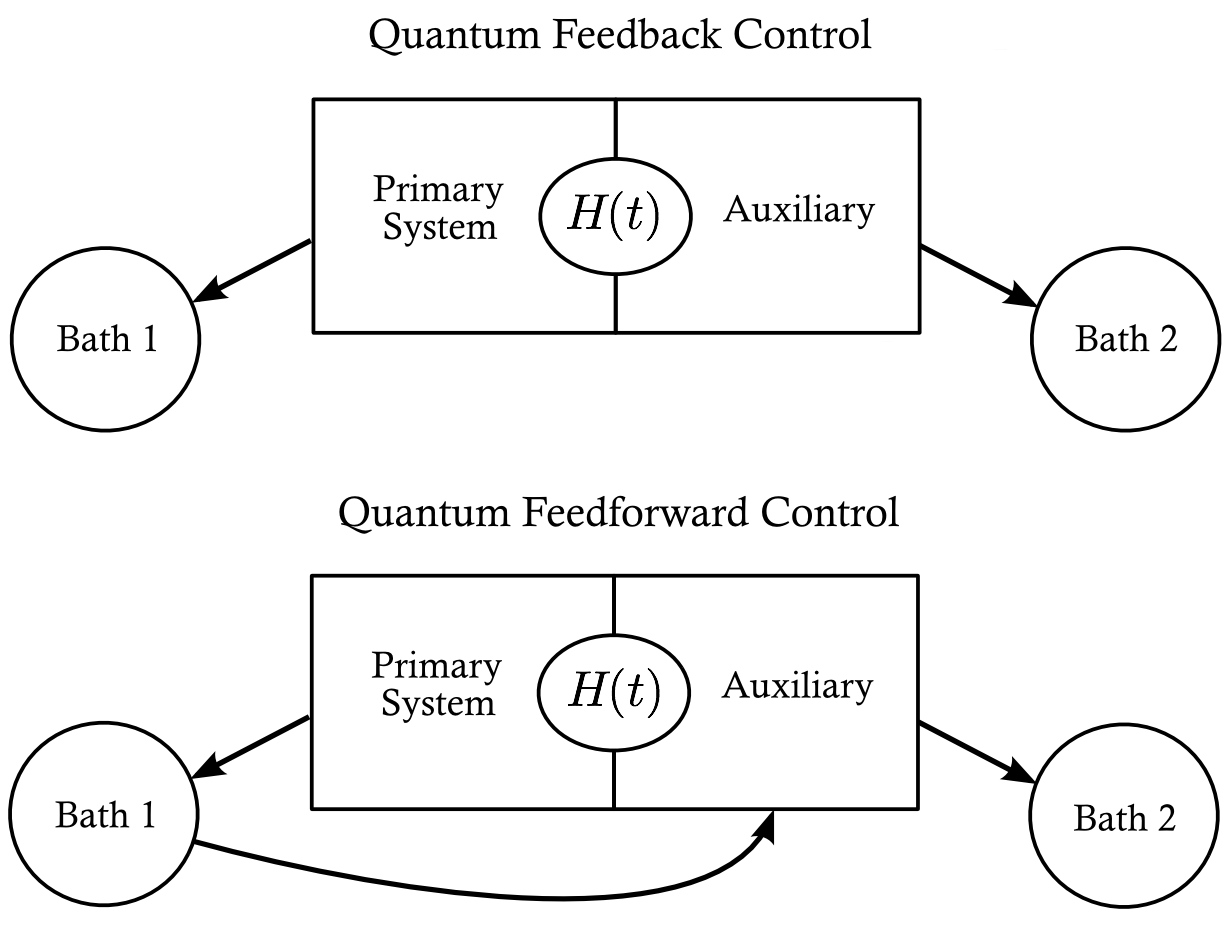}
\vspace{-1mm}
\caption{Depictions of the most general implementation of feedback (top) and feedforward (bottom) control of quantum systems. In the former the auxiliary system, or controller, does not have direct access to the noise driving the system. The direction of the arrows represents the direction in which the flow of information impacts the control problem: In feedback control neither bath provides any useful information to the systems, while in feedforward the auxiliary can extract useful information from bath 1 about the noise it injects into the system.} 
\label{fig1} 
\end{figure}

To distinguish between the two ways of implementing feedback control we refer to the one that involves explicit measurements as \textit{measurement-based feedback control}, and the one without measurements as \textit{coherent feedback control}. Let us now consider the most general form of feedback control and see how the different categories of control appear within this form. Since one can always subsume measurements into unitary evolution by including an auxiliary system, the most general feedback control scenario need not refer to measurements. We can therefore define this control in the following way. A primary system $S$ is coupled to an auxiliary system $C$ (here $C$ stands for \textit{controller}). Both the system and controller will in general also be coupled to environments (other quantum systems) that inject noise into them, and we denote these respectively by $\mathcal{E}_S$ and $\mathcal{E}_C$. The system and auxiliary evolve under a time-dependent Hamiltonian $H(t)$, whose purpose is solely to control the evolution of $S$. In addition to designing some or all of the time-dependence of $H$, we may also have some limited control over the coupling of each of the two systems to their environments. This is the most general form of feedback control, and is depicted in Fig.~\ref{fig1}(top). 

If the controller $C$ is a quantum system whose states are not directly accessible to an observer, then the above scenario is coherent feedback control. If some of the states of the auxiliary, or more generally \textit{subspaces} of the auxiliary, are macroscopically distinct and accessible to an observer, then the scenario is that of measurement-based feedback control. In this case the auxiliary must have within it a means of amplification. That is, the microscopic auxiliary states that become correlated with the microscopic states of the system must be mapped to macroscopically distinct states, whose energy  is necessarily much larger than that of the microscopic states. 

Shortly after posting the first version of this monograph, I had the opportunity to discuss the above definitions of feedback and feedforward with Daniel Lidar, Robert Kosut, Edmond Jonckhere, and Herschel Rabitz. Kosut pointed out that there was more than one definition of feedforward in the classical control community, and further, Jonkhere mentioned that historically some categories of control have been very hard to precisely define. One well-known control theorist at a famous discussion of ``adaptive'' control at an IEEE meeting remarked that ``the only way to know if a control protocol was adaptive was to ask the designer''. It was also suggested that clarifying the role of Markovian vs. non-Markovian baths in these definitions would be useful. In the Appendix I discuss these points in more detail, elucidating the different classical notions of feedback and feedforward. It may be that grey areas between these categories are inevitable, and it may also be important not to be overly zealous in asserting a single definition as superior. Nevertheless it is important to understand the motivation and justification for any definition, and to specify clearly the definition used in a given piece of work. Below I  explain why feedback can be defined without reference to a feedback ``loop''. I thank Francesco Ticozzi for hospitality at a meeting in Padova, and Robin Blume-Kohout and John Gough for discussions at that meeting that were very helpful in clarifying why loops are redundant in the definition of feedback. 

To introduce some terminology, we note that the purpose, or \textit{objective}, of a control procedure may be to prepare a particular state at a given time (state-preparation), to realize a particular unitary evolution for the system over some time interval, or to have the system evolve in some specified way as a function of time. A specific choice for the auxiliary system and the joint time-dependent Hamiltonian, designed to achieve a specific objective is called a control \textit{protocol}. 

There are a few points that are worth noting about the above scenarios, and we pause to discuss these now. 

\subsection{Everything within open-loop control}

If we remove the auxiliary system, or equivalently decide that we wish to control the whole of the joint system, $J = S \otimes A$, then the feedback control scenario reduces to that of open-loop control, traditionally referred to simply as \textit{quantum control}, in which the Hamiltonian is varied as function time to achieve an objective. In a sense feedback control is open-loop control in which we wish ``only'' to control a subsystem of the system to which we apply the control. 

\subsection{Where is the feedback loop?}

Traditionally feedback control involves extracting information from a system, and then using this information to determine forces (controls) that are applied to the system. The applied controls are often referred to as being \textit{conditional}, or \textit{conditioned}, on the measurement results. The flow of information to the controller via the measurements, and then back to the system via the control forces, is described as a \textit{loop}. This is why feedback control is also referred to as \textit{closed-loop control}. In the description of feedback control depicted in Fig.~\ref{fig1} we have dropped the loop, and this is something that tends to confuse control theorists. We now explain why the existence of a loop in feedback control is merely a conceptual device; one can just a well describe feedback without using a loop. 

The simplest, and possibly the clearest way to explain the redundancy of a loop structure in feedback control is to use an example, and ours will be the Watt governor. This is a device for controlling the speed of a steam engine. The control device, the ``governor'', was a metal ball that would swing outward under the centrifugal force of the circular motion of the engine. The ball was attached to a valve that controlled the flow of fuel to the engine. As the speed of the engine increased, the ball would swing further out, and this would reduce the flow of fuel. The governor thus provided a dynamics that stabilized the speed of the engine. We also see that that the engine and the governor are merely two dynamical systems coupled together: the equations of motion of the engine have terms that depend on the state of the governor, and those of the governor depend on the state of the engine. This relationship is symmetric, in that there is no causal ordering of the effects of each system on the other. There is nothing that says that the effect of the system on the controller is one that provides information to the controller, and that the effect of the controller on the system is based on this information. There is merely an interaction between the two systems. A Hamiltonian formulation of the dynamics of the two systems would have one or more interaction Hamiltonians, but would not involve any ``loop''. Thus the ``loop'' in feedback control is merely a conceptual tool. 

We can speak of a Hamiltonian interaction between two systems as implementing feedback control because the controller has no access to the noise driving the system. Thus anything that the controller does to reduce the effects of this noise must exploit information that it obtains from the system. While there is no loop in the description of the feedback, and while there may not be an obvious way to quantify the flow of information, especially for quantum systems, the action of the controller falls within the natural definition of feedback control. 

\subsection{Wither feedforward control?} 

Feedforward control, as defined in classical control theory, is a scenario in which the controller has access to the noise that is affecting the system before it reaches the system. The controller can then determine forces to apply to the system based on this noise and then \textit{feed these forward} to the system to counteract the noise. This is not usually considered for quantum systems because one does not usually have useful  access to the environment. But there could be situations in which feedforward is useful for controlling quantum systems, and we depict this situation in Fig.~\ref{fig1}(bottom). The only difference between feedback and feedforward is that in the latter the auxiliary system has access to the environment $\mathcal{E}_S$.  

\subsection{Reversible or irreversible coupling?} 

In measurement-based feedback control one often considers measurements in which information is continually extracted from the system, so called ``continuous measurements''~\cite{JacobsSteck06}. Continuous measurements are realized by coupling the system to an environment, or bath, in which the interaction is irreversible and continuously carries off information about the system. The archetypical example of this is the act of measuring the position of an object by shining a beam of light on it and observing the reflected light. The many modes of the traveling-wave electromagnetic field (the light) are a bath with which the system interacts, and the reflected light continually carries away information.   

If instead of measuring the reflected light in the above example, we injected this light into an auxiliary quantum system, and then allowed the auxiliary to act in some way upon the system, we would have replaced the continuous measurement-based feedback control with a version of coherent feedback control. The information extraction by the controller is mediated in this case by a quantum electromagnetic field. We could similarly apply the feedback forces by coupling the auxiliary to the system by a quantum field, where in this case the field travels from the auxiliary to the system. This scenario explicitly involves information traveling in one direction around a loop. While this puts it explicitly in the form of a classical feedback loop, it necessarily involves a loss of information via the irreversible connection to the fields. In fact, field-mediated coupling, often called a $\textit{cascade}$ connection because of the one-way flow of information~\cite{Gardiner93, Carmichael93}, is equivalent to a direct (reversible) coupling between the system and auxiliary, with the addition of irreversible coupling of each to separate environments. As a result coherent control via cascade connections is a strict subset of coherent feedback that involves direct coupling as depicted in Fig.~\ref{fig1} (top). 

\vspace{-3mm}
\section{Coherent vs. Measurement-based feedback control}

Techniques for controlling quantum systems via their interaction with another system have been studied and devised for some decades, and these are all ultimately examples of coherent feedback control. Nevertheless many general questions about coherent feedback control are as yet unanswered, as well as many regarding the relationships between coherent feedback control (CFC) and measurement-based control (MBC). First there are two simple practical differences between CFC and MBC that are worth noting. In MBC, since the information is amplified and transformed to classical numbers, this information can be processes by a classical computer. This processing is therefore noise-free and can be rather sophisticated. The disadvantage is that the amplification process, because it must be achieved with very low noise, is technologically challenging and can be slow. The merits of CFC is exactly the converse. Since it requires non measurement or amplification, it can be fast and is technologically simpler, since it only involves coherent manipulation of quantum systems via externally applied classical fields. The flip side is that the controller is now an auxiliary quantum system, which will in general be subjected to noise and dissipation from its environment.   Further, the simpler the auxiliary system, the less processing it will be able to do on the information that it obtains from the system. 

\begin{table}[t]
\addtolength\tabcolsep{2pt}
\begin{center}
\begin{tabular}{c|c}
\hline
 Meas.-based feedback (MBC) & Coherent feedback (CFC)  \\  \hline 
  noise-free processing    & controller has noise/damping  \\  
  requires amplification  &   fast, no amplification  \\  \hline  
\end{tabular}
\caption{\label{ch1::tab1} The relative practical merits of CFC and MBC}
\vspace{-8mm}
\end{center}
\end{table}

\subsection{Quantifying and comparing dynamical resources}

The first problems we consider involve the relationship between CFC and MBC. There is  a difference between a reversible coupling, such as the unitary interaction between two microscopic systems, and the irreversible coupling between a system and bath, such as a quantum field. This difference is that in the reversible coupling, so long as the energies of the two systems are bounded, if the systems start in a product state, the evolution of the entanglement generated between the systems is initially second order in time. This is enforced by the unitarity of the evolution. This behavior is distinct from the evolution generated by the interaction with a bath, at least in the Markovian approximation, because the evolution of the entanglement, and the populations of the states of the system is first-order in time. This means that the dynamical resources provided by reversible and irreversible coupling are distinct in some sense. 

An important question in any kind of control is how well an objective can be realized given the available resources, and thus the classification of resources is important. When we have at our disposal a unitary interaction between a system and auxiliary, a natural resource to consider is the difference between the maximum and minimum eigenvalues of the interaction Hamiltonian. This quantity is bounded in practice, and bounds the rate at which we can extract information from the system, which in turn bounds the rate at which we can extract entropy and reduce the effects of noise~\cite{WVSJ13}. If we instead have an irreversible coupling to a zero-temperature bath, the natural resource is the size of the damping rate induced by this coupling. It is not clear, however, whether there is a way to compare the eigenvalue bound for unitary interactions to the damping rate of an irreversible coupling. These may be comparable in some theoretically or practically motivated way, or they may be distinct. 

\vspace{1mm}
\textbf{Problem 1:} Can reversible and irreversible couplings be compared quantitatively, and thus be regarded as manifestations of a single resource for quantum control, or do they constitute distinct resources? 
\vspace{1mm}

Feedback control via continuous measurements is an example of an irreversible Markovian coupling between the primary system and the auxiliary. The efficacy of field-mediated control can therefore be compared directly with that of feedback control via continuous measurements. But a comparison between the latter and coherent feedback control implemented solely with direct (unitary) coupling between the two is only meaningful if unitary and irreversible couplings can be compared.  

Some light might be shed on the above question by examining the relationship between field-mediated and direct coherent control. As mentioned above, field mediated control is equivalent to a direct coupling combined with an irreversible bath coupling for both the system and auxiliary. From this it is clear that control via direct coupling can be compared to continuous measurement-based control if we allow the former to be augmented by coupling the system and auxiliary to appropriate baths.  

\vspace{1mm}
\textbf{Problem 2:} What is the precise relationship between continuous measurement-based feedback, and coherent feedback via direct system-auxiliary coupling given the same dissipative resources? Is the latter superior to the former, and if so, does this remain true without the dissipative resources? 

\vspace{1mm}
\textbf{Problem 3:}  Coherent feedback mediated by fields encompasses continuous MBC, except for the processing power of a classical computer. In the former the measurement results must be processed in some way via the evolution of the auxiliary. Determine the size and dynamics required of a quantum auxiliary to approximate various kinds of classical information processing of the measurement record.  
\vspace{1mm}


\subsection{Feedback control via unitary interactions}

For coherent feedback control in which the system interacts with the auxiliary via a Hamiltonian whose maximum eigenvalue difference, $\Delta\lambda \equiv \lambda_{\ms{max}} - \lambda_{\ms{min}}$ is bounded, it has been shown that coherent feedback control can extract entropy from the system  faster than measurement-based feedback under the same bound. This means that CFC will outperform MBC under this bound for a wide range of objectives. To understand why this is, we need to understand more precisely the difference between coherent feedback and measurement-based feedback. 

We stated above that all MBC protocols can be reproduced by CFC protocols, but the reverse is not true: MBC protocols are a subset of CFC. To see this we note first that all CFC operations on the primary system, since they involve the interaction with an auxiliary which is then traced out, can be written in the Krauss form 
\begin{equation}
   \rho(t) = \sum_n A_n \rho(0)  A^\dagger_n . 
\end{equation}
for an arbitrary set of operators $\{ A_n \}$ that satisfy $\sum_{A_n^\dagger A_n = I}$. A general efficient measurement on the primary is also of this form, where each measurement result, labelled by $n$, places the system in the state $\rho_n = A_n \rho A^\dagger_n/\mathcal{N}$ (the normalization is $\mathcal{N}$). We can then perform feedback, which is a unitary operation on the system for each result $n$. Averaging over the measurement and feedback, the final state of the system is 
\begin{equation}
   \rho(t) = \sum_n U_n A_n \rho(0)  A^\dagger_n U^\dagger_n . 
\end{equation}
The MBC form looks just a general as the CFC form, but there is a catch. In order that that MBC reduce the entropy of the system, and actually utilize the feedback step (if it didn't it wouldn't be MBC) there must be more than one measurement outcome, because a single unitary preserves the (von Neumann) entropy. What is shown in~\cite{Jacobs13} is that, under a bound on the Hamiltonian that performs the interaction between the system and auxiliary (an interaction that is also required to make a measure), the fastest way to extract all the entropy from the system using CFC is to follow a geodesic in the joint Hilbert space of the system and auxiliary~\cite{Jacobs13}. The operation on the system that results from this geodesic has only a single Krauss operator: 
\begin{equation}
   \rho(t)_{\ms{geo}} = |\psi\rangle \langle \psi | \rho(0) |\psi\rangle \langle \psi |  , 
\end{equation}
where $|\psi\rangle$ is the final pure state in which the protocol leaves the system. Because of this an MBC protocol cannot follow a geodesic, and must take longer to extract all the entropy.  

The result in~\cite{Jacobs13} provides a greater understanding of the relationship between CFC and MBC, and provides a guiding principle for the design of CFC protocols. It suggests that interactions that follow geodesic evolution provide superior performance for noise reduction not only for reversible but also irreversible coupling. Nevertheless...  

\vspace{1mm}
\textbf{Problem 4:} Is it possible to show quantitatively how the restriction to measurement and feedback limits the performance of feedback via irreversible coupling, and how this relates to geodesics? 
 
\vspace{1mm}
\textbf{Problem 5:} What are the geodesics that correspond to specific control objectives, and how closely can these be approximated by interactions available in mesoscopic systems?  

\vspace{1mm}
\textbf{Problem 6:} Coherent feedback via unitary interactions is inherently a periodic process. What kind of Hamiltonians are required to obtain continuous or quasi-continuous control, and is it possible to obtain simple analytic expressions for steady-state performance in some regimes? 

\subsection{Limits to quantum control}  

A practically important and fundamentally interesting question is how well various control objectives can be achieved given specific resources. When controlling mesoscopic systems, a major experimental constraint is the strength, or speed, with which we can interact with the system. The problem with obtaining theoretical limits to feedback control is that to do so we must find optimal protocols. For measurement-based control this is near-impossible, since the equations of motion are both stochastic and nonlinear. But there may be a route to finding upper bounds by exploiting the connection between coherent and measurement-based control. In general, while the equations for coherent feedback control are linear, since they involve only the Schr\"{o}dinger equation, they are just as complex as the nonlinear equations of MBC. However the problem simplifies if we consider the regime of good control. Good control implies that the control protocol is able to achieve the objective with a fidelity close to unity. This means that the rate at which it is able to extract entropy from the system is much faster than the rate at which the noise generates entropy, and thus the speed of the interaction with the system is correspondingly fast. Fortunately the regime of good control is of practical importance, and its simplifying feature is that the noise can be treated as a perturbation on the dynamics. 

In~\cite{WVSJ13} the authors were able to obtain the optimal (reversibly coupled) coherent feedback protocol for preparing pure states (e.g. for cooling) in the presence of thermal noise, in the regime of good control. The authors assumed an ideal auxiliary system so as to obtain the limits to cooling under a constraint on the speed of the interaction with the system. This shows that it is possible to obtain optimal results for CFC in the regime of good control, and it may therefore be possible to use similar results to obtain bounds on measurement-based control. 

In~\cite{WVSJ13} the authors did not solve the most general problem of cooling in the regime of good control, but restricted themselves to the regime in which the interaction with the system does not change its energy eigenvalues appreciably. The reason for this restriction is that it corresponds to the regime in which the control operations do not change the dynamics induced by the thermal noise. The Markovian Redfield master equation that describes this noise depends on the system's energy levels. Thus the authors of~\cite{WVSJ13} solve the problem of optimal CFC for the case in which the master equation is fixed. Relaxing this restriction would provide the answer to the fundamental question of just how well one can prepare pure states given a maximal speed of interaction with the system, taking account of the fact that one can significantly alter the dynamics of the system during the process. While the structure of the more general problem is somewhat different to that solved in~\cite{WVSJ13}, the results in~\cite{WVSJ13} suggest that it may be tractable. Further, for a Markovian and therefore memoryless bath, it is easy to determine the Master equation as a function of the control Hamiltonian: the Master equation must immediately adapt to the new Hamiltonian at each point in time. Thus the master equation is merely the Redfield equation implied by the Hamiltonian of the joint system at each time. 

\vspace{1mm}
\textbf{Problem 7:} Determine the optimal protocol for state preparation via a unitary interaction with a system, in the regime of good control, under a bound on the norm of the interaction Hamiltonian, and a bound on the rate at which the Hamiltonian can be changed. 
\vspace{1mm}

Often there is a practical bound on the interaction strength between two mesoscopic systems, but this does not stop us from coupling a system of interest to multiple auxiliary systems. As we increase the number of auxiliary systems, the effective coupling rate between the system and the set of auxiliaries increases.  

\vspace{1mm}
\textbf{Problem 8:} Determine the optimal protocol for state preparation via a unitary interaction with the system, in the regime of good control, under a bound on the norm of the interaction Hamiltonian with a given auxiliary, as a function of the number of auxiliary systems. 
\vspace{1mm} 

We mentioned above that it may be possible to bound the performance of continuous MBC by using bounds obtained on CFC, where the CFC is chosen to have equivalent resources (irreversible coupling strength). The reason for this is that CFC subsumes MBC. We also mentioned that reversibly coupled CFC is equivalent to irreversibly coupled CFC with the addition of coupling to specifically chosen baths. Thus by finding optimal solutions for reversibly coupled CFC (as in problems 6 and 7), but with the addition of some irreversible bath coupling, it may be possible to obtain useful bounds on continuous MBC. 

\vspace{1mm}
\textbf{Problem 9:} Determine bounds on the performance of continuous measurement-based feedback control, in the regime of good control, by finding optimal protocols for coherent feedback control using equivalent resources. 

\subsection{Linear field-mediated feedback control}

A subset of CFC problems that is being investigated by a number of groups from the control theory community is that of linear quantum systems coupled (irreversibly) via quantum fields. These systems are also referred to as quantum linear networks. It is well-known that linear systems have a very restricted dynamics, in that they cannot, for example, create non-Gaussian quantum states from Gaussian ones. It is also known that linear networks cannot perform universal quantum computation in an efficient manner. Nevertheless there are other tasks that one would expect that linear networks cannot perform, but proofs have not yet be obtained.  

\vspace{1mm}
\textbf{Problem 10:} Show that linear networks cannot perform quantum error-correction. 
\vspace{1mm} 
 
Error-correction involves correcting errors on arbitrary states, that is states about which the error correcting process is ignorant. A task that is simpler is protecting known states from the effects of noise and decoherence. 

\vspace{1mm}
\textbf{Problem 11:} Show that linear networks cannot protect non-Gaussian states from linear noise. 
\vspace{1mm} 

So far work on linear feedback networks has been focussed on time-independent networks, in which the Hamiltonians of the systems and the couplings between then do not change with time. The analysis in~\cite{WVSJ13} suggests that including time-dependence in networks can increase their performance for the purposes of control. 
 
\vspace{1mm}
\textbf{Problem 12:} Quantify the increase in performance provided by including time-dependent external control in linear feedback networks, and examine how this varies across classes of problems.  
\vspace{1mm} 

We mentioned above that direct (reversible) coupling between systems is more flexible than irreversible (field-mediated) coupling, because the latter comes with bath coupling. This means that a larger class of linear networks can be realized by using direct coupling. 

\vspace{1mm}
\textbf{Problem 13:} Quantify the increase in performance that direct coupling provides over field-mediated coupling, with or without external time-dependent control. 
\vspace{1mm} 

\subsection{Imperfect controllers} 

So far we have considered feedback control as if the auxiliary system were perfect. For measurement based control we can take the processing of the measurement results to be perfect but not necessarily the actions of the controller on the system. For coherent feedback control we cannot even assume that the controller itself, and thus the processing of the information is noiseless. In fact, if the system that we wish to control is noisy, often the auxiliary system will be just as noisy. If this were not the case then why not use the auxiliary in place of the system for that ever task we have in mind? 

Sometimes there is a physical justification for why the controller may be different from the system. One example of this is ``resolved-sideband'' cooling, a simple example of CFC in which a mechanical resonator is cooled by coupling it to an electrical or optical auxiliary resonator. In this case the electrical/optical resonator has such a high frequency that it has essentially zero entropy at the ambient temperature. It is this fact that allows the interaction to transfer entropy out of the mechanical resonator to the auxiliary, cooling the former. But the auxiliary, while it has zero entropy and is in this sense less noisy, still undergoes damping. While this damping is essential to dump the extracted entropy into a reservoir, and thus to obtain steady-state cooling, the analysis in~\cite{Wang11} shows that this damping reduces the ability of the control to prepare the purest possible state at a single time (this state is more pure than the best achievable steady-state). 

\vspace{1mm}
\textbf{Problem 14:} Determine optimal control protocols in the regime of good control when the auxiliary system is subject to significant damping. 
 
\vspace{1mm}
\textbf{Problem 15:} What are the limits to control when the controller is just as noisy as the system? How does this limit scale as the size of the auxiliary is increased in relation to that of the system?  
 
\vspace{1mm} 
\textbf{Problem 16:} How does noise on the control Hamiltonian $H(t)$ affect the fidelity? Can noise on the auxiliary and noise on the Hamiltonian be managed in the same way? 

\vspace{1mm}
\textbf{Problem 17:} Compare the performance of coherent and measurement-based control protocols in which the quantum controller is damped and/or noisy. At what level of noise is the advantage provided by coherent feedback eliminated by the noise, and is this important for real applications? 
\vspace{1mm}

\section{The complexity of open-loop control}

We now turn to open-loop control, in which there is no auxiliary system, and the task is simply to change the Hamiltonian with time so as to provide a desired evolution. While I have used the word ``simply'' open-loop control is in fact highly complex. Usually one is faced with a Hamiltonian of the form $H = \sum_n c_n H_n$, where the classical parameters $c_n$ can be varied with time in a more-or-less arbitrary way. The problem of constructing a given unitary $U$ in some time $T$ by choosing the $c_n(t)$ is in general tremendously complex and has no analytic solution. One way to determine the required $c_n(t)$ is to use numerical search methods. To generate a unitary with dimension $N$, given only a fixed number of parameters $c_n$, the complexity of the control functions $c_n(t)$ will in general scale as $N^2$, since the number of free parameters in a unitary of size $N$ has this scaling. Nevertheless, given certain kinds of Hamiltonians $H_n$, not all unitaries are equally difficult to produce. In quantum computing, for example, the set of Hamiltonians is two-body interactions between qubits. Those unitaries that can be generated with a sequence of two-body unitaries the length of which scales polynomially with the number of qubits are referred to as fast algorithms (those that scale exponentially are slow).  

In the introduction we explained how open-loop control is intimately related to feedback control, since the latter is merely open-loop control on the system and auxiliary. Questions about the difficulty of performing open-loop control therefore have a direct bearing on feedback control. The resources for quantum control are a set of available Hamiltonians $H_n$ in the joint space of the system and auxiliary. A problem is easy if the complexity of the time-dependent control scales as a polynomial in the logarithm of the size of the system 

\vspace{1mm}
\textbf{Problem 18:}  For what control objectives, and what set of control Hamiltonians, is  open-loop control easy? Does the set of easy objectives depend in a non-trivial way on the set of control Hamiltonians? 
\vspace{1mm}

\section{Miscellaneous problems}

I promised 20 open problems and so far I have only delivered 18. Here are two more problems to satisfy the title. 

\vspace{1mm}
\textbf{Problem 19:} Continuous-measurement based feedback control is highly complex because modifying the measurement as part of the feedback can increase the performance. Optimal protocols are still not known even for a single qubit, and progress is hampered by the size of the search space. Explore more complex protocols for small quantum systems than those that have been tried to date to see if they enhance performance, guided by known results to-date. For protocols that do enhance performance, consider whether this enhancement has a bearing on coherent feedback. 
 
\vspace{1mm}
\textbf{Problem 20:} Are there situations in which feedforward control would be useful in mesoscopic quantum systems? If so explore this paradigm of control with explicit examples and compare it to feedback control. 
\vspace{1mm}

\section{Addendum} 

Quantum control lies at the intersection of information, measurement, and unitary dynamics, and is interesting for this reason. It also has an impact on other fields for which precise control over quantum systems is required, and thus a wide range of potential applications. 
By painting a broad picture of quantum control, and the complexity and range of open questions in this field, I hope that I have stimulated your interest. With luck you might even feel motivated to investigate some of these questions...
 
\textit{Acknowledgements:} The author is partially supported by the NSF projects PHY-1005571 and PHY-1212413, and by the ARO MURI grant W911NF-11-1-0268.

\section{Appendix: Two definitions of feedback and feedforward} 
\label{appB}

\begin{figure}[t]
\leavevmode\includegraphics[width=0.98\hsize]{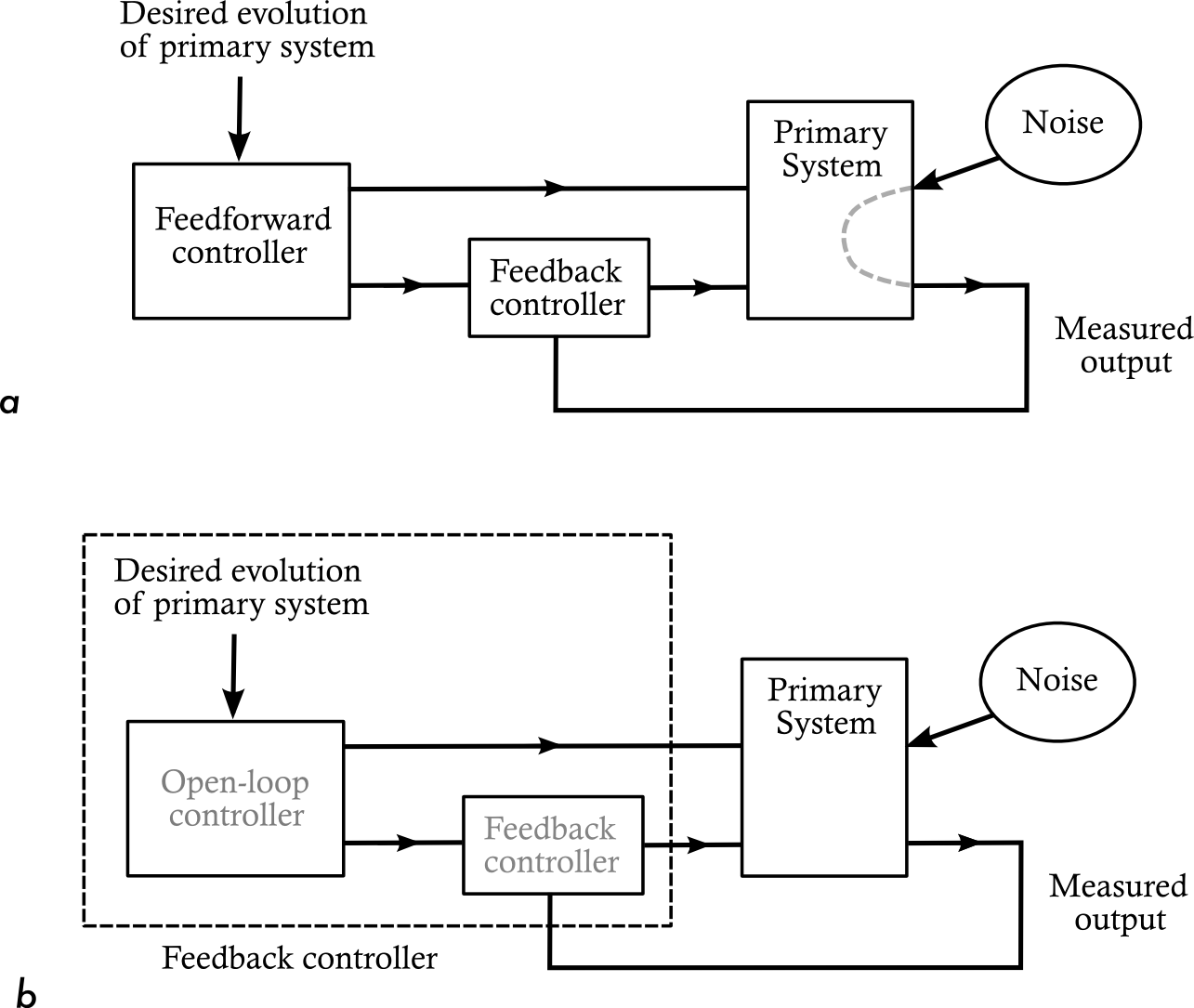}
\vspace{-1mm}
\caption{Top: An alternative classical definition of feedback and feedforward. Bottom: The classical definition of feedback that we use.} 
\label{fig3} 
\end{figure}

In Fig.~\ref{fig3}a we display an alternative definition of feedforward in classical control systems. We see from the diagram that the box denoting the ``feedforward'' controller has no access to information from the system, and in our language is therefore referred to as an open-loop controller. Further, in Fig.~\ref{fig3}a the feedback controller can be allowed access to the noise realization driving the system at the same time as it effects the system (as indicated by the dashed grey line), but does not have access to the noise \textit{prior} to its effect on the system. It is clear that Fig.~\ref{fig3}a is a quite different nomenclature for defining classes of control than the one we use. In Fig.~\ref{fig3}b we show how our nomenclature would be used to describe the controller configuration in Fig.~\ref{fig3}a. Everything inside the black dashed line is the feedback controller, and we have removed the grey dashed line. 

Under our terminology a controller cannot perform feedforward unless it has some access to the noise realization driving the system over and above its interaction with the system. But we have to decide whether or not to include in this definition the situation in which the controller has access to the noise realization at the present time, or to restrict it solely to the case when it has advance knowledge of this noise (that is, \textit{prior} to the time at which it affects the system). To keep our definitions simple, we choose here to define feedback control as the situation in which the controller can interact with the degrees of freedom of the system only, and has no access to any additional information regarding the noise realization driving the system. It may of course have information about the parameters that characterize the noise, such as the variance(s), correlation function(s), etc. We correspondingly define feedforward control as any situation in which the controller receives information about the noise realization without interacting with the system.  

\subsection{Markovian vs. non-Markovian baths}

The only reason that we include separate baths for the primary and auxiliary in Fig.~\ref{fig1} is to make it clear the auxiliary is unable to extract useful information from the bath that couples to the primary. If the bath's are Markovian, or more importantly if they are \textit{effectively Markovian}, meaning that the memory time of the bath is shorter than the timescale on which we can modify the joint Hamiltonian, then even when both systems are coupled to the same bath the auxiliary is usually not able to use the bath to extract useful information. Thus in many settings in which there is only a single bath, the situation is that of feedback control, not feedforward (under our nomenclature). 

A second question concerns whether it is necessary for bath 1 to be non-Markovian in order to perform feedforward. Under our definition above, the auxiliary is a feedforward controller if it can determine the noise realization driving the system at the current time, without interacting with the system. This is possible if bath 1 is Markovian, and the auxiliary is fed the results of measurements made on it. As an explicit example, if the primary system is a single particle, and the bath induces Markovian dynamics given by the Lindblad operator $x$ (position), then the effect of the bath is that of a linear random force with a white noise spectrum acting on the particle. If the bath is measured in the right way (in the right basis), then the measurements reveal nothing about the state of the particle, but do reveal the values of the fluctuating force~\cite{doh99}, which can then be corrected for by the auxiliary. 


\begin{thebibliography}{6}%
\makeatletter
\providecommand \@ifxundefined [1]{%
 \@ifx{#1\undefined}
}%
\providecommand \@ifnum [1]{%
 \ifnum #1\expandafter \@firstoftwo
 \else \expandafter \@secondoftwo
 \fi
}%
\providecommand \@ifx [1]{%
 \ifx #1\expandafter \@firstoftwo
 \else \expandafter \@secondoftwo
 \fi
}%
\providecommand \natexlab [1]{#1}%
\providecommand \enquote  [1]{``#1''}%
\providecommand \bibnamefont  [1]{#1}%
\providecommand \bibfnamefont [1]{#1}%
\providecommand \citenamefont [1]{#1}%
\providecommand \href@noop [0]{\@secondoftwo}%
\providecommand \href [0]{\begingroup \@sanitize@url \@href}%
\providecommand \@href[1]{\@@startlink{#1}\@@href}%
\providecommand \@@href[1]{\endgroup#1\@@endlink}%
\providecommand \@sanitize@url [0]{\catcode `\\12\catcode `\$12\catcode
  `\&12\catcode `\#12\catcode `\^12\catcode `\_12\catcode `\%12\relax}%
\providecommand \@@startlink[1]{}%
\providecommand \@@endlink[0]{}%
\providecommand \url  [0]{\begingroup\@sanitize@url \@url }%
\providecommand \@url [1]{\endgroup\@href {#1}{\urlprefix }}%
\providecommand \urlprefix  [0]{URL }%
\providecommand \Eprint [0]{\href }%
\providecommand \doibase [0]{http://dx.doi.org/}%
\providecommand \selectlanguage [0]{\@gobble}%
\providecommand \bibinfo  [0]{\@secondoftwo}%
\providecommand \bibfield  [0]{\@secondoftwo}%
\providecommand \translation [1]{[#1]}%
\providecommand \BibitemOpen [0]{}%
\providecommand \bibitemStop [0]{}%
\providecommand \bibitemNoStop [0]{.\EOS\space}%
\providecommand \EOS [0]{\spacefactor3000\relax}%
\providecommand \BibitemShut  [1]{\csname bibitem#1\endcsname}%
\let\auto@bib@innerbib\@empty
\bibitem{Note} Much less inadequate reference list to come... 
\bibitem [{\citenamefont {Jacobs}\ and\ \citenamefont
  {Steck}(2006)}]{JacobsSteck06}%
  \BibitemOpen
  \bibfield  {author} {\bibinfo {author} {\bibfnamefont {K.}~\bibnamefont
  {Jacobs}}\ and\ \bibinfo {author} {\bibfnamefont {D.~A.}\ \bibnamefont
  {Steck}},\ }\href@noop {} {\bibfield  {journal} {\bibinfo  {journal}
  {Contemp. Phys.}\ }\textbf {\bibinfo {volume} {47}},\ \bibinfo {pages} {279}
  (\bibinfo {year} {2006})}\BibitemShut {NoStop}%
\bibitem [{\citenamefont {Gardiner}(1993)}]{Gardiner93}%
  \BibitemOpen
  \bibfield  {author} {\bibinfo {author} {\bibfnamefont {C.~W.}\ \bibnamefont
  {Gardiner}},\ }\href {\doibase 10.1103/PhysRevLett.70.2269} {\bibfield
  {journal} {\bibinfo  {journal} {Phys. Rev. Lett.}\ }\textbf {\bibinfo
  {volume} {70}},\ \bibinfo {pages} {2269} (\bibinfo {year}
  {1993})}\BibitemShut {NoStop}%
\bibitem [{\citenamefont {Carmichael}(1993)}]{Carmichael93}%
  \BibitemOpen
  \bibfield  {author} {\bibinfo {author} {\bibfnamefont {H.~J.}\ \bibnamefont
  {Carmichael}},\ }\href {\doibase 10.1103/PhysRevLett.70.2273} {\bibfield
  {journal} {\bibinfo  {journal} {Phys. Rev. Lett.}\ }\textbf {\bibinfo
  {volume} {70}},\ \bibinfo {pages} {2273} (\bibinfo {year}
  {1993})}\BibitemShut {NoStop}%
\bibitem [{\citenamefont {Wang}\ \emph {et~al.}(2013)\citenamefont {Wang},
  \citenamefont {Vinjanampathy}, \citenamefont {Strauch},\ and\ \citenamefont
  {Jacobs}}]{WVSJ13}%
  \BibitemOpen
  \bibfield  {author} {\bibinfo {author} {\bibfnamefont {X.}~\bibnamefont
  {Wang}}, \bibinfo {author} {\bibfnamefont {S.}~\bibnamefont {Vinjanampathy}},
  \bibinfo {author} {\bibfnamefont {F.~W.}\ \bibnamefont {Strauch}}, \ and\
  \bibinfo {author} {\bibfnamefont {K.}~\bibnamefont {Jacobs}},\ }\href@noop {}
  {\bibfield  {journal} {\bibinfo  {journal} {PRL, in press}\ } (\bibinfo
  {year} {2013})}\BibitemShut {NoStop}%
\bibitem [{\citenamefont {Jacobs}\ and\ \citenamefont {Wang}(2013)}]{Jacobs13}%
  \BibitemOpen
  \bibfield  {author} {\bibinfo {author} {\bibfnamefont {K.}~\bibnamefont
  {Jacobs}}\ and\ \bibinfo {author} {\bibfnamefont {X.}~\bibnamefont {Wang}},\
  }\href@noop {} {\bibfield  {journal} {\bibinfo  {journal} {Eprint:
  arXiv:1211.1724}\ } (\bibinfo {year} {2013})}\BibitemShut {NoStop}%
\bibitem [{\citenamefont {Wang}\ \emph {et~al.}(2011)\citenamefont {Wang},
  \citenamefont {Vinjanampathy}, \citenamefont {Strauch},\ and\ \citenamefont
  {Jacobs}}]{Wang11}%
  \BibitemOpen
  \bibfield  {author} {\bibinfo {author} {\bibfnamefont {X.}~\bibnamefont
  {Wang}}, \bibinfo {author} {\bibfnamefont {S.}~\bibnamefont {Vinjanampathy}},
  \bibinfo {author} {\bibfnamefont {F.~W.}\ \bibnamefont {Strauch}}, \ and\
  \bibinfo {author} {\bibfnamefont {K.}~\bibnamefont {Jacobs}},\ }\href@noop {}
  {\bibfield  {journal} {\bibinfo  {journal} {Phys. Rev. Lett.}\ }\textbf
  {\bibinfo {volume} {107}},\ \bibinfo {pages} {177204} (\bibinfo {year}
  {2011})}\BibitemShut {NoStop}%
  \bibitem{doh99} A. C. Doherty and Tan, S. M. and Parkins, A. S. and Walls, D. F., Phys. Rev. A, \textbf{60}, 2380 (1999).  
\end{thebibliography}

%

\end{document}